# Spiral Galaxies' Disk Dominance and Testing MOND

G.G. Byrd (1) and S. Howard (2)

(1) U. Alabama and (2) USNO /retired.

Many Hubble type S galaxies have a flat rotation curve that extends into the outer disk. The surface brightness there is too low to create the curve from stars of a reasonable mass-to-light ratio. To maintain a stable disk one usually assumes a massive dark matter halo that dominates the rotation. Gravitational arm amplification is used here to estimate if the curve is created by a 100% disk, a mixture, or by a 100% halo. The disk surface density is not estimated from the disk surface brightness and mass-to-light ratio. Instead, the long-term maximally amplified arms' pitch angle is used to estimate disk surface density. The rotation curve is proportional to the square root of surface density. Surprisingly, a loose arm galaxy disk can dominate in creating the rotation curve. For example, NGC 3198's arm pitch angle of 30 degrees implies a disk surface density which accounts for 88% of the rotation curve, a dominant disk. Little dark matter halo is needed for such galaxies. Observationally, the dark outer disk may be molecular hydrogen. NGC 3198's dominant disk also provides a counter example to the need for any universal (MOND) force law modification over the outer disk spiral arms. In contrast tightly wound flat curve galaxies are not estimated to have dominant disks, e.g. NGC 7217's 4.8 degree arm pitch angle only accounts for 41% of its rotation curve.

## Introduction

Many Hubble type S galaxies have flat rotation curves (constant circular orbital speed, $V$, versus radius, $r$) extending into their outer disks (Whitehurst and Roberts 1972, Roberts and Whitehurst 1975, Rubin *et al* 1978, Bosma 1978). The surface brightness in the outer disk is too low to create $V$ from stars with a reasonable mass-to-light ratio. For stabilization the disk dynamics are usually considered to be dominated by dark matter halos, particularly to create the flat $V$ versus $r$ (Byrd and Valtonen 2020 review).

Disk perturbations can evolve into a conventional density wave inside and outside a strongly amplified co-rotation radius (Lindblad 1960),

(Linn and Shu 1966, and Lin, Yuan, and Shu 1969). Byrd and Howard (2021) evaluated the disk surface densities of a set of type S galaxies assuming a tightly wound density wave. Unexpected disk dominance results were found in this tightly wound approximation. A more realistic general evolutionary loose arm approach is thus carried out in the present paper.

An analytic method is used here to estimate whether a flat rotation curve of a grand design spiral galaxy is created by a 100% disk, something in between, or by a 100% halo. This paper uses the arm winding in S type galaxies to estimate the outer disk surface mass density, $\mu$, and the amount it contributes to the flat $V$ compared to the halo. This method does *not* use the disk surface brightness or any assumed disk mass-to-light ratio. The approximate study of Byrd and Howard (2021) is explored more realistically here.by not assuming the arms are tightly wound. Computer simulations serve as an independent check of the validity of these analytic estimates.

## Arm Properties and Definitions

The arm radius is taken to increase with increasing disk azimuthal angle, $\varphi$, which increases opposite the orbital motion. The angular rate ($\Omega = V/r$) is taken to decline with $r$. The arm multiplicity is $m = 2$ for an arm pair *etc*. This paper derives $\mu$ as a function of $r$ using the orbital speed $V$, and $\gamma=90°-i$ which is the acute angle between the extended arm and the extended radius. The spiral arm pitch, $i$. is the acute angle between the extended arm and the orbital arc at the crossing point.

Danver (1942) and Kennicutt (1981) find observationally that S galaxies have logarithmic (log) arms in which the pitch (tan $i$) is constant as the arm $r$ increases. Two-armed components are predominant (Considere and Athanassoula 1988). Log arms have a constant arm tan $i$, or, alternatively, a constant arm cos $\gamma$. The log arm equations are:

$$\varphi = \varphi_0 - \frac{p}{m}\ln(r) \qquad (1)$$

$$\cot i = \frac{p}{m} = -r\frac{d\varphi}{dr} \qquad (2)$$

Hubble type *S* galaxies show a range of arm pitches. NGC 7217 has a tightly wound arm pair with, $i = 4.8°$ (Buta *et al*. 1995, Figure 10) and a flat *V* (Kalnajs 1983). Elmegreen *et al*. (1989) fit log spirals to arm patterns and obtained 16°, 18°, and 15° respectively in the disks of M81 (NGC3031), M100 (NGC4321) and M51 (NGC5194). Honig and Reid, (2015, Figure 5) examined HII regions in M101 (NGC4321) to trace a two- arm pattern whose $i = 20°$. By going steeper NGC 3198's arms have $i = 30.0 \pm 6.7°$ (Ferrarese 2002). All have nearly flat rotation curves over their outer arm regions.

The normal wavelength $\lambda_n$ is the perpendicular distance crest to crest. The wave number, *k*, is given by:

$$|k| = \frac{2\pi}{\lambda_n} = \frac{2\pi}{(2\pi r/m)\cos\gamma} = \frac{m}{r\cos\gamma} \qquad (3)$$

### Arm Properties Compared to Computer Simulations

As shown in Figure 1, M51 has log arms in the main part of the disk. (Howard and Byrd 1990 or Byrd 1995). Figure 2 is a log *r* versus angle plot which shows more clearly the log arms as straight lines of the dust lanes and the simulation disk particles which agree. The outermost arms show irregularities due to the most recent second passage of the companion. Elmegreen *et al*. (1989)'s observational log plots of the arms of M51 and other galaxies are particularly striking.

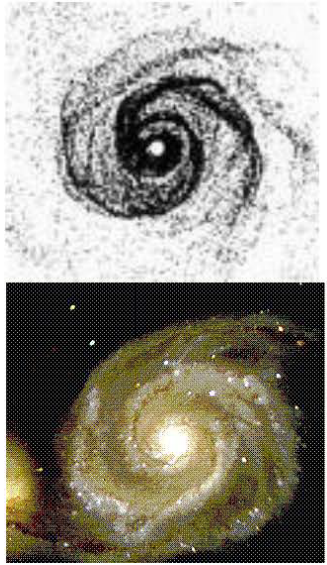

Figure 1. The top frame shows a simulation of a perturbed disk for M51 whose surface density contributes 63% of the flat *V* rotation curve. (Howard and Byrd 1990, Figure 11). Compare it to the lower photo of M51 (courtesy William Keel, University of Alabama). The simulation indicates that the companion on the lower left of the image perturbed the disk previously to trigger the disk arms.

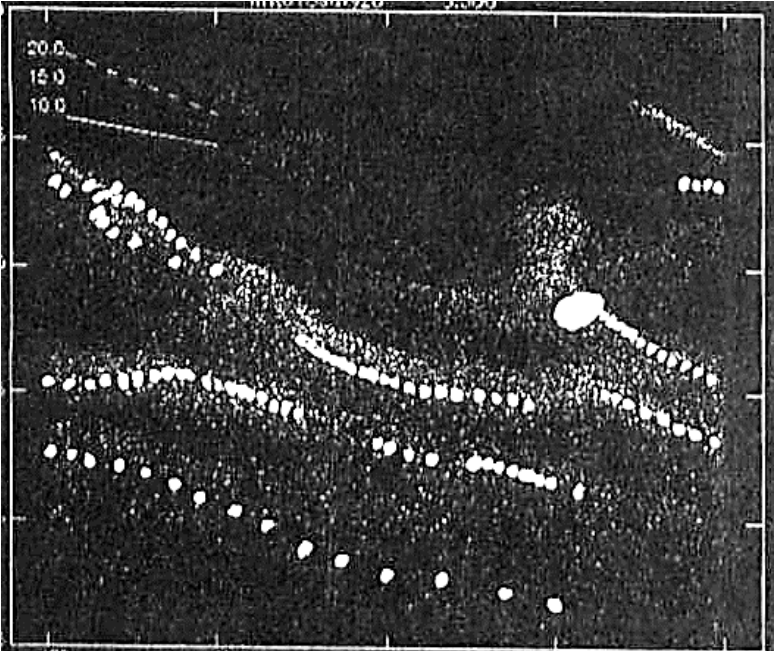

Figure 2. Display of log r versus azimuthal angle φ. Figure 1 M51 simulation particles are small dots. The first two tic marks on the vertical axis up from the horizontal axis correspond to 0.1 and 1.0 times the original disk radius. The horizontal axis tics are at 90° intervals of φ. The observed dust lanes are the larger white dots. The companion is the largest irregular ellipse. Upper left lines mark 10°, 15° and 20° slopes of straight log arms (Byrd 1995).

## Perturbation, Amplification, and Surface Density

The general case of both loosely and tightly winding arms is considered here. An initial near and far side impulsive tidal velocity perturbation is created by a passing galaxy or other source as sketched in Figure 3. Byrd and Howard (1992) find that there are very frequent low mass passing disturbers which can create tidal arms which develop over the entire disk.

The Figure 3 perturbation winds up along with epicyclic oscillation to create swing amplified arms as schematically diagrammed in Figure 4.

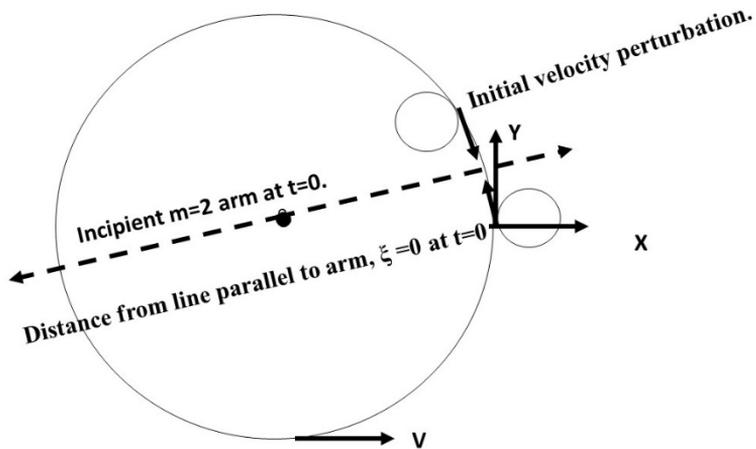

**Figure 3.** Schematic view of initial disk particles' zero positions and velocity perturbations toward an approximately straight incipient arm. The circular velocity, *V*, is counterclockwise. Epicyclic motion executes spring oscillation about a line through its initial position parallel to the winding arm.

Figure 4 shows a schematic view of the Figure 3 arm at a later time. "Opposite side" disk particles have oscillated toward one another, creating the arm density enhancement. Note the velocity perturbations parallel to the arms moving inward or outward on opposite sides of the arms. The upper left epicycles show how velocity "wiggles" across the arm or even inward and outward streaming parallel to the arms may arise. These are observed in M51 (Shetty *et al* 2007) and in our simulations.

The perturbed particles each oscillate about a line parallel to the winding arm which also passes through each particle's initial unperturbed circular orbiting x y frame shown in Figures 3 and 4. The oscillator equation is

$$\frac{d^2\xi(\gamma)}{d^2 t} + S\xi(\gamma) = 0. \tag{4}$$

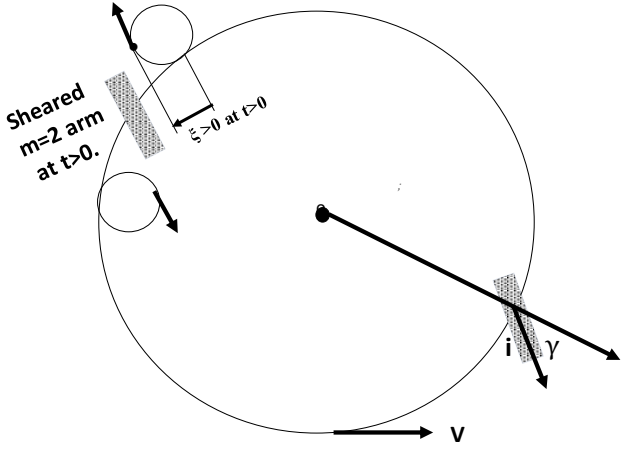

**Figure 4.** Schematic diagram of epicyclic" perturbation toward a wound-up Fig. 3 arm plus mirror image displacement on other side. ξ is the distance between the line parallel to the arm through the epicycle center and a line parallel to the arm through the particle.

The following discussion is phrased in terms of a harmonic oscillator (equation 4). The dispersion relation parameter $S$ is a "spring constant" for oscillation relative to the winding arm as described in Toomre (1981). $S$ provides a condition for stability determined by the competing disk density self-gravity, $\mu$, the stabilizing dispersion, $c$, orbital angular rate, $\Omega = V/r$, and shear. The impulsive arm wave tends to wind up from an m-fold ~radial $\gamma \approx 0°$. We use Toomre's (1981, equation 20) version of the Goldreich/Lynden-Bell (1965) dispersion relation which is suited for the disk. The dispersion relation is:

$$S = \kappa^2 - 8\Omega A \cos^2 \gamma + 12 A^2 \cos^4 \gamma - 2\pi G k \mu + k^2 c^2 \qquad (5)$$

Aside from variables already defined, $c^2$ is the velocity dispersion squared. The two "$A$" terms describe the arm looseness:

$$A = -\frac{r}{2}\frac{d\Omega}{dr} = \left(\frac{1}{2}\frac{V}{r}\right)_{Vconst} . \qquad (6)$$

See Michikoshi and Kokubo (2016) and Toomre (1981) for more information. The right bracket subscript in equation (6) and henceforth has a constant *V*. The tightly wound epicyclic frequency curve is given by equation (7a) below both for a general or a flat rotation curve *V*. Equation (7b) gives the epicyclic frequency for generalized loose or tightly wound arms with the subscript. *L*. Substitute equation (6) for A in equation (7b) to get Equation (7c) which describes generalized arm winding and flat rotation curve with *L,V* subscripts .

$$\kappa^2 = r^2 \frac{d}{dr}\left(r^4 \Omega^2\right) = 4\Omega^2 + 2r\Omega\frac{d\Omega}{dr} = \left[\frac{2V^2}{r^2}\right]_{V=\text{const}}. \quad (7a)$$

$$\kappa_L^2 = \kappa^2 - 8\Omega A \cos^2 \gamma + 12 A^2 \cos^4 \gamma \quad (7b)$$

$$\kappa_{L,V}^2 = 2\left(\frac{V}{r}\right)^2 \left[1 - 2\cos^2 \gamma + \frac{3}{2}\cos^4 \gamma\right] \quad (7c)$$

Continuing the evolution of Figures 3 and 4, as time increases and the arm winds, γ increases from approximately radial (~0º) toward tightly wound (approaching 90º). Consequently, cos γ smoothly decreases with time from ~1. During this process the arm is amplified from zero to a longer-term maximum strength. This would be appropriate for the strong-armed grand design spirals. A time sequence n-body simulation of a perturbed disk and the amplification to long-term log arm is shown in Figure 5. There is only one disturber passage, so the arms are more regular in appearance than for M51. The amplified long-term arm pattern arm pitch is much like that in the M51 Figure 1. This is because the simulation shown in Figure 5 has the same disk contribution to the total rotation curve as that of Figure 1. The rest is contributed by a an inert "dark halo.

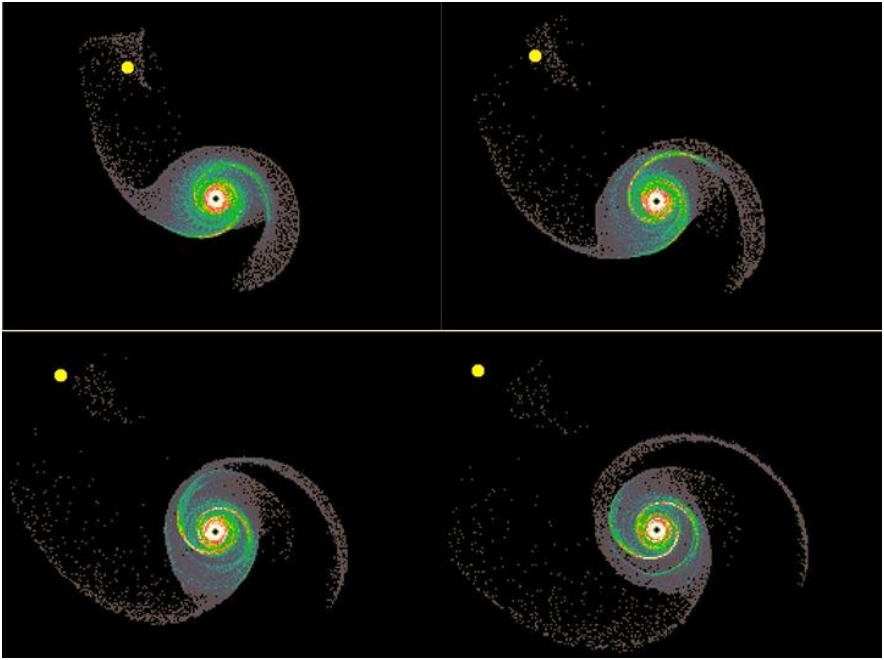

Figure 5. Amplification arm evolution of a tidally perturbed disk

## Tightly Wound Arms and Stability Dispersion

Because it is commonly used in discussions (*e.g.* in Chap. 6 in Binney and Tremaine 2008, and Chap. 12 in Shu 1982), the disk surface density will first be calculated assuming tightly wound arms in equation (5) for *S=0*. In other words, cos (γ) is small so the "*A*" terms in equation (5) can be neglected. The dispersion $c^2$ is taken to be non-zero. The tightly wound version of equation (5) is

$$S = 0 = \kappa^2 - 2\pi Gk\mu + k^2 c^2$$

Substituting equation (3) in equation (8a) to solve for c,

$$S = (\kappa - kc)^2 = (\kappa - \pi Gk\mu/\kappa)^2 = 0 \qquad (8b)$$

$$c = Q\frac{\pi G\mu}{\kappa} \qquad (9)$$

Equation (9) agrees well with the frequently used Toomre stellar disk stability criterion (Toomre 1964) with a Q of one. This is a widely used convenient criterion for the tightly wound case (subscript *T*). This dispersion is independent of $\gamma$.

In the tightly wound *S*=0 equation (8b) above, the right-side parenthesis set to zero gives the disk surface density for stability,

$$\mu_T = \frac{\kappa^2 r}{\pi G m} \cos\gamma \tag{10a}$$

Donner and Thomasson (1994) and Thomasson and Donner (2002) calculated galaxy disk surface densities with an equation like (10a) but only half as large *i.e.*, the tightly wound equation (8) with *c=0*. In this paper a two times larger density is given by equation (10a) which includes a stabilizing velocity dispersion. Equation 10a) is confirmed by the results at co-rotation of Shu (1982). For a flat rotation curve with *V* the tightly wound (*T,V*) disk surface density ($\mu_{TV}$) is

$$\mu_{TV} = \left(\frac{4\cos\gamma}{m}\right)\left(\frac{V^2}{2\pi G r}\right)_{100\%M} = F_{TV}\mu_{100\%V} \tag{10b}$$

where the second parenthesis in (10b) is the 100% disk surface density ($\mu_{100\%V}$) of a flat rotation curve Mestel (1963) disk. The first parenthesis is the fraction ($F_{TV}$) of the 100% disk indicated by the maximally amplified arm winding.

Results very similar to equations (9), (10a), and (10b) can also be obtained with a force law softening which provides a close analog to a stellar velocity dispersion reduction factor as described in Byrd (1995), Binney and Tremaine (2008), and Miller (1971).

### Amplified Loose Arms and Collective Relaxation

In this section the tightly wound approximation will be removed and reveal significantly different results. As noted earlier, equation (9) agrees well with the frequently used Toomre stability criterion (Toomre 1964) with a Q of one. Observationally, does Q typically differ from 1 and, if so, by how much?  Binney and Tremaine (2008) provide information to estimate a typical Q from the local gas and stars. The "cold" gas apparently dominates with Q=1.5.  This can be taken as an estimate for the factor that a typical disk might exceed the Toomre stability tightly wound criterion.

How does $c$ arise in galaxies? How can the observed Q be greater than one? There are a variety of possibilities. Mechanisms for the origin of the velocity dispersion are discussed by Binney and Tremaine (2008). One proposed mechanism is an increase in dispersion via gravitational encounters with molecular clouds which Binney and Tremaine believe to be an unlikely primary cause of disk heating.  Another mechanism is the gravitational effects of spiral features. Binney and Tremaine describe that a grand design density wave with a fixed pattern speed will not ``heat" the stellar disk. They conclude that another mechanism, transitory chaotic spiral arm heating can explain the origin of the dispersion.  Here we describe a winding spiral arm heating mechanism which will create a dispersion which will specifically stabilize the disk.

We will see how a difference from a "tightly wound" Q value greater than one can arise by a more general approach.  For loose or tight arms, the full dispersion relation equation (5) applies but with the more general $\kappa$ (Lynden-Bell 1965, Toomre 1981).

$$S_L = 0 = k^2 c^2 - 2\pi G k \mu + \kappa_L^2 \qquad (11)$$

$\kappa_L$ is given by equation (7b). Using equation (7c) assuming a constant $V$, we obtain

$$S_{LV} = \frac{2V^2}{r^2}(1 - 2\cos^2\gamma + \frac{3}{2}\cos^4\gamma) - 2\pi G \mu m / (r\cos\ \gamma) + (\frac{m}{r\cos\gamma})^2 c^2 \qquad (12)$$

As was done for the tightly wound approximation, we solve for the stabilizing disk velocity dispersion and surface density for loosely as well as tightly wound arms. The loosely wound stabilizing dispersion, $c_L$, can be obtained from equation (11) as follows.

$$0 = \kappa_L^2 - 2\pi Gk\mu + k^2c^2 = (\kappa_L - kc)^2 = (\kappa_L - \pi Gk\mu/\kappa_L)^2$$

$$c_L = \frac{\pi G\mu}{\kappa_L} = Q\frac{\pi G\mu}{\kappa} \qquad (13)$$

By examining equations (7) we see that $\kappa_L$ is less than $\kappa$, resulting in Q>1. Generally, Q equals one only for very tightly wound arms. Equation (13) differs from the tightly wound approximation.

    A collective relaxation process provides a comparatively quick origin of the dispersion necessary for stability and for motions normal to the disk plane. If $c$ initially is zero, then the $S$ in equation (11) will amplify and dip below zero as arms wind up from initial small perturbations. The disk surface density and critical arm winding $\gamma_c$ result in the stabilizing dispersion given by equation (13). Toomre (1981) has plots of this "dip below zero" where the harmonic oscillator becomes unstable. The dispersion will arise in a collective spiral arm relaxation process similar to that proposed in systems as large as clusters of galaxies and as small as globular clusters (Lynden-Bell 1967).

    A natural limit is thus placed on the collective process in the evolution of a given initial set of tidally induced arms. After this first critical winding arm amplification is reached and a sufficient dispersion is created, the arms will dissipate. Subsequent perturbations will not readily heat a disk which already has the stabilizing dispersion (equation (13)) from a previous triggering. These subsequent perturbations would create arms which reach maximal strength at the same critical $\gamma_c$ at which the dispersion was created then subsequently weaken.

    Figure 6 shows a simulation of the dissipation process in a perturbed disk.The top frame shows the regular maximally amplified arms. The bottom two show the unstable dissolution due to softening that simulates a velocity dispersion. This is an approximate simulation. A description is given in Byrd (1995). Also see Toomre (1981).

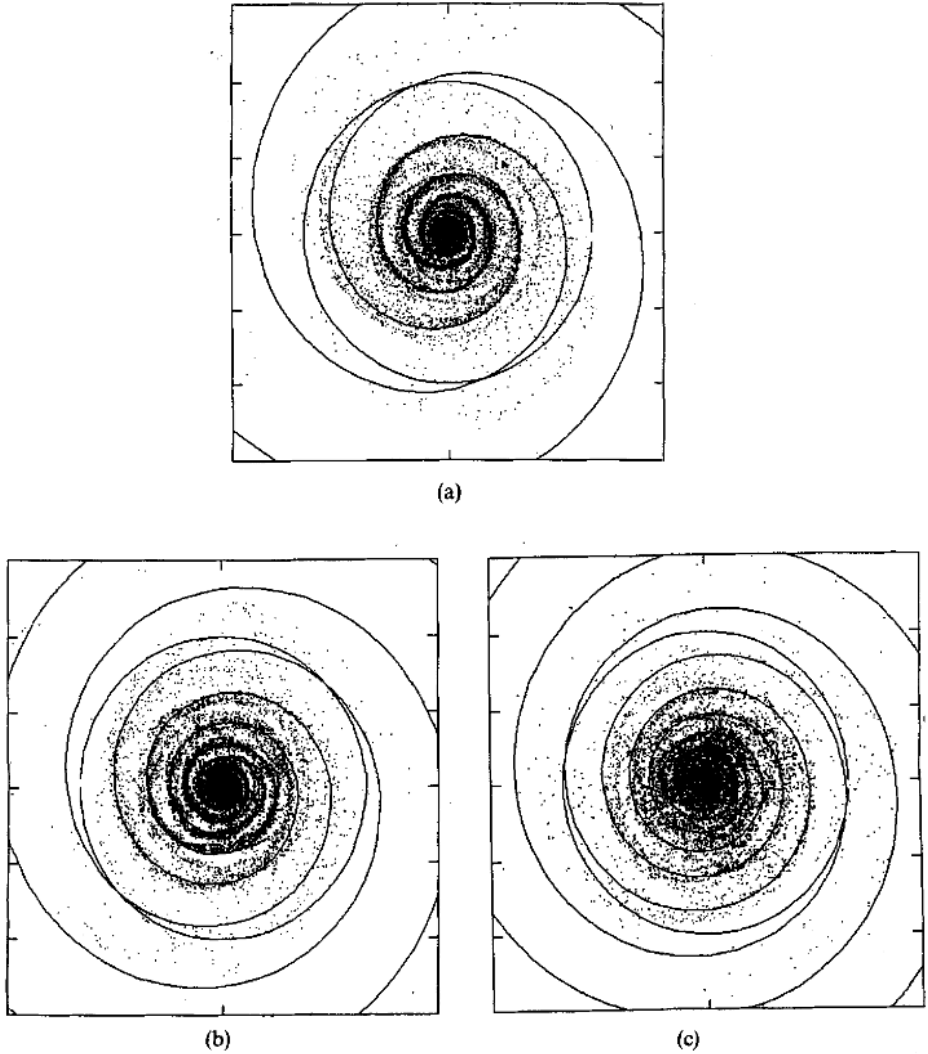

**Figure 6.** Amplification arm evolution to dispersion of a tidally perturbed disk. The circle is the initial disk radius. The spiral is a log arm fitting to the tidally triggered arms. Frame (a) is the maximally amplified arm pattern. Frames (b) and (c) show the breakup after instability is reached.

## Loose Arms and Disk Surface Density

We will now analytically obtain the relation between the disk surface density and the maximally amplified arm winding after a stabilizing disk dispersion has been generated. After a perturbation the arms will wind up to be a maximally amplified "grand design" (*i.e.* S=0) but not dip below zero (*i.e.* dS /dcos γ = 0).   See Byrd (1995) for a discussion. By using these two conditions in the following expression, we find the righthand side is zero.

$$\frac{d(S\cos^2 \gamma)}{d\cos \gamma} = \frac{dS}{d\cos \gamma}\cos^2 \gamma + S\ 2\cos \gamma = 0 + 0$$

Substituting equation (12) in the left side parenthesis and taking the derivative gives equation (14a), the surface density for maximally amplified loose or tight arms with the equation (13) velocity dispersion.   The observable variables, V, r, and γ permit calculation of the surface density, μ.

$$\mu_L = \frac{r}{\pi G m}\kappa^2 \cos\gamma(1 - 4\cos^2 \gamma + \frac{9}{2}\cos^4 \gamma) \quad (14a)$$

$$\mu_{LV} = \frac{4\cos\gamma}{m}(1 - 4\cos^2 \gamma + \frac{9}{2}\cos^4 \gamma)\left[\frac{V^2}{2\pi G r}\right]_{100\%V} = F_{LV}\mu_{100\%V} \quad (14b)$$

Equation (14b) shows the loose arm disk surface density as a fraction, $F_{LV}$, of the 100% surface density of a flat rotation curve disk inside the square brackets, $\mu_{100\%V}$. By checking we see that equation (14b) with γ approaching 90° matches equation (10 a, b) which is the disk surface density assuming tightly wound arms.

The present paper's tight or loose arm (14a,b) results  are a more general and accurate formulation than those which assume tight arms. Table I shows the fractional surface density results $F_{LV}$ of maximally amplified tight or loose arms using equation(14b) for a flat rotation curve. These are compared to the results for the tightly wound approximation.  Looking at the last three columns, the tightly wound approximation seems pretty good up to 74° winding or 16° pitch. For looser arms (steeper pitch) the errors can be larger.

Table I shows that in the tight arm approximation, a galaxy with a 60° winding or 30° pitch has a disk surface density equal to that of a 100% flat rotation curve disk. However, the more accurate loose arm formulation shows that a galaxy with arms of that pitch has a surface density of 0.78 of that of a 100% disk.  The more accurate loose arm formulation shows that a galaxy with a somewhat looser 54° arm winding or 36° arm pitch would have a 100% disk surface density.

**Table I. Two-fold Arm Winding (γ) or Pitch(i) Estimate of Flat V Disk Surface Density.** Columns (3) and (4) are disk surface density, μ., as fractions of a 100% flat $V$ disk in the loose, $F_{LV}$, and tightly wound, $F_{TV}$, (3) and (4) cases. (5) is the "error" of $F_{TV}$ relative to the loose $F_{LV}$.

---

| γ° | i°= 90-γ° | Loose Frac $F_{LV}$ V Disk μ | Tight Arm Approx Frac $F_{TV}$ Disk μ | Tight Arm μ % "error" vs Loose |
|---|---|---|---|---|
| **53.88** | **36.12** | **1.000** | **1.179** | **18%** |
| 54 | 36 | 0.995 | 1.176 | 18% |
| 55 | 35 | 0.951 | 1.147 | 21% |
| **60** | **30** | **0.781** | **1.000** | **28%** |
| 62 | 28 | 0.730 | 0.939 | 29% |
| 64 | 26 | 0.685 | 0.877 | 28% |
| 66 | 24 | 0.645 | 0.813 | 26% |
| 68 | 22 | 0.605 | 0.749 | 24% |
| 70 | 20 | 0.566 | 0.684 | 21% |
| 72 | 18 | 0.525 | 0.618 | 18% |
| 74 | 16 | 0.482 | 0.551 | 14% |
| 76 | 14 | 0.435 | 0.484 | 11% |
| 78 | 12 | 0.383 | 0.416 | 8% |
| 80 | 10 | 0.328 | 0.347 | 6% |

## Disk Contribution to Rotation Curve: Example Galaxies

The disk surface density is informative but not so easy to directly connect to observations with the disk local dispersion and vertical structure in galaxies other than our own. The disk rotation curve is more directly measured for comparison with our calculations of disk dominance.

In Table I the loose arm formulation shows that a galaxy with a 54° arm winding or 36° arm pitch angle would have a 100% disk surface density. The galaxy would have a flat rotation curve equal to *V* or $V_{observed}/V=1$ or 100%. What if the galaxy has less than 100%? We use equation (14b) to compare the observed rotation speed to that of a 100% disk. Take the square root to compare the two at the same radius,

$$\frac{V_{disk\,contrib}}{V_{100\%}} = \sqrt{\frac{\mu_{LV}}{\mu_{100\%V}}} = \sqrt{F_{LV}} = \sqrt{\frac{4\cos\gamma}{m}(1 - 4\cos^2\gamma + \frac{9}{2}\cos^4\gamma)} \qquad (15)$$

Equation (15) is applied to sample type S galaxies in Table II, column two. NGC 3198's arms have $i = 30.0 \pm 6.7°$ *i.e.* the pitch angle may be as large as $i = 37°$ (Ferrarese 2002). NGC 3198 has an extended very flat rotation curve (Begeman 1989). By examining the last right hand column top line of Table I, and using equation (15), it is clear that NGC 3198 has a dominant disk relative to any halo out to the arm limits which can account for 88% of the observed rotation curve, *V*, a dominant fraction. See Figure 7 for an image of this loose-armed galaxy. There can evidently be a considerable amount of less luminous matter within the spiral arm disk of this galaxy.

Going to the other extreme among our sample galaxies, in the fourth column last line of Table II, the 4.8° pitch of NGC 7217 indicates a disk component that can account for 41% of the observed *V* (important but not dominant). See Figure 8 for an illustration of this tightly wound galaxy and compare it to the strikingly different Figure 7 of the disk dominated NGC 3198. The "in between" galaxies in terms of dominance are also intermediate in terms of morphology.

A general evolutionary loose arm approach is carried out in the present paper. The general sense of our earlier approximate tightly wound results (Byrd and Howard 2021) is confirmed. Disk surface densities of some type S galaxies were estimated to be gravitationally dominant over any inert halo or other component even under the approximation of a tightly wound density wave.

**Table II. Arm Winding Estimate of Fraction 100% Flat V Disk Surface Density for sample Type S Galaxies**

| Galaxy NGC # Messier # | Arm winding and (Pitch), $\gamma°$; and ($i°=90-\gamma°$) | Loose Mestel D. Frac. =$F_{LV}$ m=2 | $V_{Disk\,Contrib}$ / V = $\sqrt{F_{LV}}$ |
|---|---|---|---|
| NGC3198 | 60(30) | 0.781 | 88% |
| M101 | 70(20) | 0.566 | 75% |
| M100 | 72(18) | 0.525 | 72% |
| M51 | 75(15) | 0.459 | 68% |
| NGC7217 | 85.2(4.8) | 0.165 | 41% |

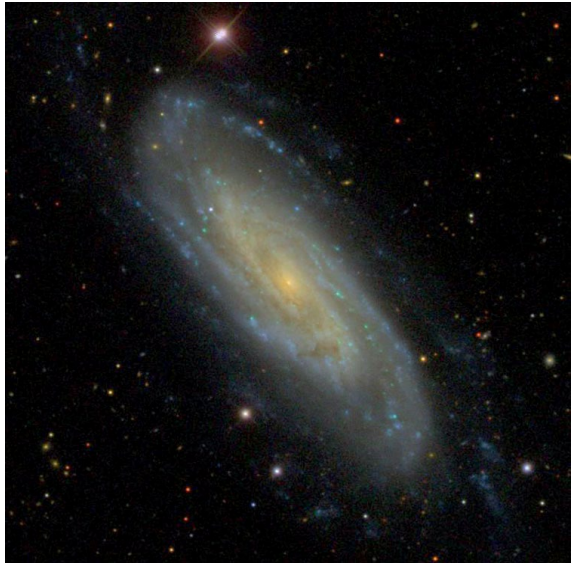

Figure 7.  NGC 3198, whose arm pair's 4.8° pitch angle indicates a disk contributing 88% of the observed rotation curve, *V*, (dominant). www.sdss.com.

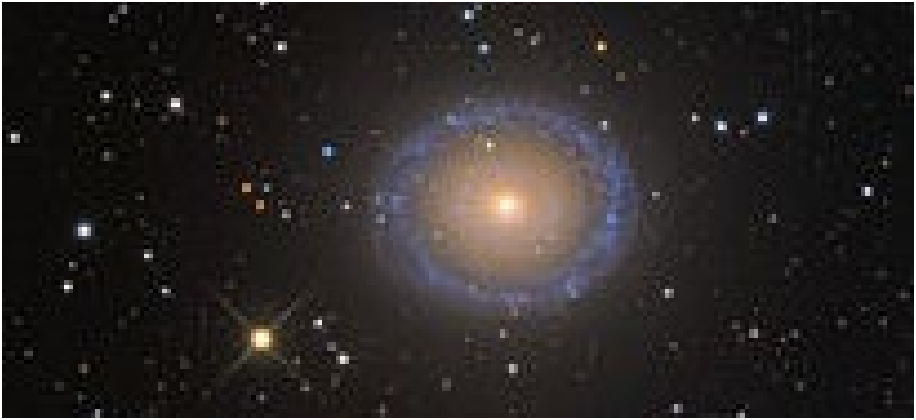

Figure 8. NGC 7217 whose arm pair's 4.8 ° pitch angle indicates a disk contributing 43% of the observed *V* (important but not dominant). www.sdss.com.

## Possible Dark Disk Substantiation

Is dark matter in a disk as dominant as that indicated for NGC 3198 reasonable? As mentioned earlier, assuming a reasonable mass-to-luminosity ratio, the stellar disk material is quite insufficient to alone create the full observed flat rotation curve in the outer disk. Recently observed dark molecular hydrogen clouds are a possibility for a majority dark disk mass. A detailed proposal has been made that molecular hydrogen is a strong baryonic candidate (Pfenniger *et al* 1994a, 1994b). Although molecular hydrogen is difficult to detect, it has now been observed in our own galaxy, its companions, and in other galaxies (Shull 2022).

A convincing example is the European Space Agency's infrared space telescope's, ISO, detection of molecular hydrogen in the edge-on spiral galaxy NGC 891 which shows the disk rotation curve in the disk. Estimating the amount of molecular hydrogen in the disk indicates that there are amounts sufficient to have all the dark matter baryonic (Valentjin 1999). Well-behaved simulations of disk dominated galaxies have been carried out (Pfenniger 2001, Reviz et al 2006).

## Gravitationally Dominating Disks and a Test of MOND

Recall that a modified law of gravity (MOND) is another explanation for the flat curves (Banik, I. and Zhao, H. 2022, review). In contrast to the dark matter halo, MOND explains the flat *V* of the outer disk by modifying the Newtonian force law on the disk material due to the galaxies' interior mass. *No dark matter halo is assumed, and the baryonic disk is judged to be insufficient to create a flat rotation curve under conventional gravitation*. In MOND, in the outer disk, the gravitation of each part of the interior matter is taken to decrease more slowly than the inverse square law ($1/r$ rather than $1/r^2$). Thus, the rotation curve remains flat as *r* increases.

It turns out that the results here can contradict the MOND hypothesis. MOND is supposed to be "universal", applying to all galaxies. A dominant disk in a galaxy such as NGC 3198 contradicts MOND theory. NGC 3198's predominant disk over the arm region is consistent with standard Newtonian gravity for the flat *V*. The need for MOND is reduced or nearly removed over the range of the arms.

## Conclusions

We showed how to calculate the surface mass density of the disks of "grand design" Hubble type S galaxies as a function of radius. See equations (14a,b). The observed degree of arm winding or pitch angle is used along with the rotation curve. The rotation curve can be non-linear or flat. Also, differing from common discussions, the arms are not necessarily assumed to be tightly wound. The disk surface density importance is obtained over the arms' extent. Arms will be conspicuous (like "beads on a string") when they reach maximal amplification from a small initial tidal perturbation. Finally, a disk dispersion naturally results in a value sufficient for stability *e.g.* the Toomre stability value for tightly wound arms but larger for looser (less wound up) cases.

The fractional disk contribution to the observed total rotation curve is obtained for the common case of nearly flat rotation curves and measured arm winding or pitch angle (Equation 15). For example, in NGC7217 whose tightly wound-up arm pitch angle is 4.8º, the disk contributes 41% of the full rotation curve over the spiral arms with the rest from a more gravitationally static dark matter halo or another component. For an extreme loose arm case, NGC 3198 has an arm pitch angle of 30°. Equation (15) indicates that NGC 3198 has a dominant disk relative to any halo out to the arm limits which can account for 88% of V, a dominant gravitating disk. Other cases are intermediate between these two values.

Some galaxies' spiral arms and rotation curves imply an extended gravitating non-luminous disk component which can create most of the flat curve. A reasonable dark baryonic candidate, molecular hydrogen, has been observationally detected. A single extreme case like NGC3198 whose disk may totally generate almost all of its flat rotation curve can call into question the need for a universal MOND gravitational force modification out to the distance limits of the spiral arms.

Our results provide a starting point for more informed investigations *e.g.,* of halo dark matter or high dispersion stars versus disk dark matter for different galaxy Hubble types. For example, it has been found observationally that there is an anti-correlation between the central black hole mass and arm inclination in spiral galaxies. (Seiger *et al*. 2008, Berrier *et al*. 2013a, b, Savchenko and Reshetnikov 2013). A physical connection of low disk mass (relative to halo) with central black holes is suggested.

## References

Banik, I. and Zhao, H, 2022 *Symmetry* **14**(7), 1331.

Begeman, K.G.1989 *Astronomy & Astrophysics,* **223**, 47B, https://ui.adsabs.harvard.edu/abs/1989

Berrier, J. C., Davis, B. L., Kennefick J., and Seigar M. S. Lacy C. H., 2013, *The Astrophysical Journal*, **769**, 132.

Berrrier, M., Meert, A., Sheth, R. K., Vikram, V., Huertas-Company, M., Mei, S., and Shankar, F. 2013, *Monthly Notices of the Royal Astronomical Society*, **436**, 697.

Binney, J. and Tremaine, S. 2008 *Galactic Dynamics* (Princeton University Press), Chap. 6.

Bosma, A. 1978. Ph.D. thesis, Univ. of Groningen. The Netherlands.

Bosma, A. 2002, "HI and dark matter in spiral galaxies," The Dynamics, Structure and History of Galaxies, ASP Conference Series, vol. 273, p. 223. G. S. Da Costa and E. M. Sadler eds.

Buta, R., van Driel, W., Braine, J., Combs, F., Wakamatsu, K., Sofue, Y., and Tomita, A, 1995 *The Astrophysical Journal*, **450**, 593.

Byrd, G. G. (1995) Proceedings of the Waves in Astrophysics conference. Editors J. H. Hunter and R. E. Wilson. *Annals of the New York Academy of Sciences*, **773,** 302. https://www.researchgate.net/publication/229817541_Tidal_Perturbations_Gravitational_Amplification_and_Galaxy_Spiral_Arms

Byrd, G. and Howard, S. 1992 *Astronomical Journal* **103**, 1089

Byrd, G.G., and S. Howard. 2021, *Journal of the Washington Academy of Sciences* **107**, number 1, 1. https://www.researchgate.net/publication/350676201_Spiral_Galaxies_When_Disks_Dominate_their_Halos_using_Arm_Pitches_and_Rotation_Curves

Byrd, G. and Valtonen, M.2020. *Astronomy & Geophysics*, **61**(6):6.26-6.29

Danver, C. G. 1942. Lund Ann., No. 10.

Donner, K. J. and Thomasson, M. 1994, *Astronomy and Astrophysics*, **290**, 785

Elmegreen, B. G., Elmegreen, D. M., and Seiden, P. E. 1989, *The Astrophysical Journal*, **343**, 602.

Ferrarese, L. and Merritt, D. 2000, *Ap. J. Letters*, **539**, L9

Goldreichh, P. and Lynden-Bell, D. 1965, *MNRAS*, 130, 97.

Honig, Z. N. and Reid, M. J. 2015, *The Astrophysical Journal* **800**, 53.

Howard, S. and Byrd, G. 1990, *The Astronomical Journal* **99**, 1798.

Kalnajs, A. 1983, *Internal Kinematics and Dynamics of Galaxies*, IAU (International Astronomical Union) Symposium 100. L. Athanassoula, editor, p. 87.

Kennicutt, R. C. Jr. 1981, *The Astronomical Journal* **86,** 1847.

Lindblad, P. O. 1960 *Stockholm Obs. Ann* **21**, 4.

Linn, C. C. and Shu, F. H. 1966. *Proc. of National Acad. of Sciences*, USA, **55**: 229

Linn, C. C, Yuan, C. and Shu, F. 1969. *Astrophysical Journal*. **155**: 721.

Lynden-Bell, D. 1967, *Monthly Notices of the Royal Astronomical Society* **136**, 101.

Mestel, L. 1963, *Monthly Notices of the Royal Astronomical Society* **126**, 553.

Michikoshi, S. and Kokubo, E. 2016  Swing Amplification of Galactic Spiral Arms: Phase Synchronization of Stellar Epicycle Motion https://arxiv.org/abs/1604.02987v1

Miller, R. H. 1971, *Astrophysics and Space Science* **14**, 73.

Pfenniger, D. 2001, "Maximum Galactic Disks versus Hot Dark Halos," Gas and Galaxy Evolution PASP Conference Series, eds. J. E. Hibbard, M. P. Rupen and J. H. van Gorkom.  Vol **240**, 319.

Pfenniger, D, Combes, F. and Martinet, L. 1994a, *Astronomy & Astrophysics,* **284**, 79;

Pfenniger, D. and Combes, F. 1994b *, Astronomy and Astrophysics* **285**, 94

Raffikov, R. R. 2001, *Monthly Notices of the Royal Astronomical Society*, **323**, 445.

Reviz, Y., Pfenniger, D., Combes, F. and Bournaud, F, 2006*, Astronomy and Astrophysics* **501**, 171.

Roberts, M.S. & Whitehurst, R.N. 1975 *The Astrophysical Journal*   **201**, 327

Rubin, V. C., Kent, Jr. W., and Thonnard, N. 1978 *The Astrophysical Journal. J.* **225**, L107.

Safronov, V. S. 1960, *Astronomy and Astrophysics*, **23**, 979.

Savchenko, S. S. and Reshetnikov, V. P.  2013, *Monthly Notices of the Royal Astronomical Society*, **436**, 1074.

Seigar M. S., Kennefick D., Kennefick J., Lacy C. H., 2008, *The Astrophysical Journal Letters* **678** (2): L93–L96.

Shetty, R., Vogel, S. N., Ostriker, E. C., and Teuben, P. J. 2007 *The Astrophysical Journal,* **665**:1138.

Shull, J. M. 2022, *Physics Today* **75**, 12, 12

Shu, F. H. 1982 *The Physical Universe*, University Science Books. Mill Valley California, pp. 280-281.

Thomasson, M. and Donner, K. J. 2002 Astronomical Society of the Pacific Conference Series **275**, p. 311.

Toomre, A. 1964 *Astrophysical Journal* **13**, 1217.

Toomre, A. 1981. In *Structure and Evolution of Normal Galaxies*, S. M. Fall and D. Lynden-Bell, Eds.: p. 1137. Cambridge Univ. Press. Cambridge, England.

Valentjin, E. A. 1999 *Ap.J.* 522, L29

Whitehurst, R.N. & Roberts, M.S.1972 *The Astrophysical Journal* **175**, 347